\documentclass[12pt]{iopart}
\usepackage[]{qcircuit}
\expandafter\let\csname equation*\endcsname\relax
\expandafter\let\csname endequation*\endcsname\relax
\usepackage{amsmath, amssymb}
\usepackage{mathtools}
\usepackage{verbatim}
\usepackage{blkarray}
\usepackage{array,makecell}
\usepackage{mathrsfs}
\usepackage{braket}
\usepackage{float}
\usepackage{listings}
\usepackage{xcolor}

\newcommand{\grayScale}{0.95} 
\definecolor{codeBackground}{rgb}{\grayScale ,\grayScale ,\grayScale}
\definecolor{forestGreen}{rgb}{0.13,0.55,0.13}

\usepackage[
style=numeric-comp,
sorting=none
]{biblatex}
\lstnewenvironment{codelisting}{
    \lstset{
    language=C++,
    backgroundcolor=\color{codeBackground},
    tabsize=4,
    showstringspaces=false,
    showtabs=false,
    showspaces=false,
    basicstyle=\ttfamily,
    identifierstyle=\ttfamily,
    keywordstyle=\color{blue},
    stringstyle=\color{red},
    commentstyle=\color{gray},
    numberstyle=\color{magenta},
    morecomment=[l][\color{forestGreen}]{\#}
    }
}{}

\begin{document}

\title{QuantumSkynet: A High-Dimensional Quantum Computing Simulator}

\author{Andres Giraldo-Carvajal$^1$, Daniel A. Duque-Ramirez$^1$, \\Jose A. Jaramillo-Villegas$^{1,2}$}

\address{$^1$ Universidad Tecnológica de Pereira, Pereira, Risaralda, Colombia\\
$^2$ Laboratory for Research in Complex Systems, Menlo Park, California, USA}
\ead{jjv@utp.edu.co}
\vspace{10pt}
\begin{indented}
\item[]Septiembre 2020
\end{indented}

\begin{abstract}
The use of classical computers to simulate quantum computing has been successful in aiding the study of quantum algorithms and circuits that are too complex to examine analytically. Current implementations of quantum computing simulators are limited to two-level quantum systems. Recent advances in high-dimensional quantum computing systems have demonstrated the viability of working with multi-level superposition and entanglement. These advances allow an agile increase in the number of dimensions of the system while maintaining quantum entanglement, achieving higher encoding of information and making quantum algorithms less vulnerable to decoherence and computational errors. In this paper, we introduce QuantumSkynet, a novel high-dimensional cloud-based quantum computing simulator. This platform allows simulations of qudit-based quantum algorithms. We also propose a unified generalization of high-dimensional quantum gates, which are available for simulations in QuantumSkynet. Finally, we report simulations and their results for qudit-based versions of the Deutsch--Jozsa and quantum phase estimation algorithms using QuantumSkynet.
\end{abstract}

\section{Introduction}

Quantum computing has made significant progress in recent years. The race to build a computer capable of solving problems that classical computers cannot is reaching an intense level \cite{piattini2020quantum, inside2020gideon}.

Cryptography \cite{boneh2013secure, marshall2016continuous}, drug design and discovery \cite{cao2018potential, darwish2019chemometrics, zhou2010quantum}, the study of complex molecular structures \cite{cao2019quantum, xia2020qubit}, and optimization and financial forecasting \cite{orus2019quantum, orus2019forecasting} are some areas that hope to benefit from the construction of a quantum computer. However, today's quantum computers are highly susceptible to noise, which can lead to errors in the calculations they perform \cite{nachman2020unfolding}.

An emerging alternative that allows increased resistance to noise is high-dimensional quantum computing \cite{erhard2018twisted}. Working with quantum states that handle multiple dimensions not only enables systems that are more robust to noise but also more secure systems with greater information capacity \cite{cozzolino2019high}. 

Different physical platforms can be used to implement high-dimensional quantum systems \cite{wang2020qudits}, such as photons \cite{islam2017provably, imany201850, lu2020quantum, erhard2018twisted, imany2019high, lukens2019quantum}, trapped ions \cite{low2020practical, bramman2019measuring}, superconducting systems \cite{kiktenko2015single, tan2021experimental}, and molecules \cite{sawant2020ultracold, wernsdorfer2019synthetic, moreno2018molecular}. The practical application of these qudit-based systems is a very active research field with great potential.

One tool that is instrumental in the progress in this area is quantum computing simulators. With these simulators, it is possible to study quantum circuits and measure their performance in the presence of noise, which helps make decisions that can improve the design of quantum hardware \cite{guerreschi2020intel}.

Currently, different quantum computing simulators and frameworks, such  as qHiPSTER \cite{guerreschi2020intel}, Quantum Composer \cite{zaman2021quantum}, Strawberry Fields \cite{killoran2019strawberry}, Rigetti Computing \cite{sete2016functional}, Libquantum \cite{butscherlibquantum}, and Quirk \cite{quirk}, among others, allow the execution of quantum algorithms and circuits to be efficiently simulated based on qubits; however, there is currently no quantum computing tool that allows simulation for high-dimensional quantum circuits.

In this paper, we propose QuantumSkynet, the first high-dimensional quantum computing simulator, which will enable scientists to evaluate, analyze, and tune high-dimensional quantum algorithms in a cloud-based environment. Additionally, we propose a generalization for high-dimensional quantum gates, where we show a relationship between Weyl's adjoint operator and high-dimensional Pauli gates. Furthermore, we propose a more generalized equation to describe high-dimensional Fourier quantum transform.

\section{High-dimensional quantum gates}

During the research and development of QuantumSkynet, a compendium and generalization of quantum gates were employed to build arbitrary high-dimensional quantum algorithms, as described in the following.

\subsection{Weyl operators}

Weyl operators are a type of gate that apply to a single qudit. They can be used to create states of maximum entanglement \cite{bertlmann2008bloch} \cite{baumgartner2007special}. These unitary operators form an orthonormal basis of a $d$-dimensional Hilbert--Schmidt space and are represented by $d^2$ matrices generated from the following expression:

\begin{equation}
    W_{p,q} = \sum_{k=0}^{d-1} \omega^{k\hspace{0.5pt}p}\Ket{k}\Bra{k \oplus q}
\end{equation}

where $k \oplus q$ is equal to $(k + q) \, mod \, d$, $\omega$ and is defined by $e^{ (\frac{2\pi}{d}) i }$, and $p,q$ are subscripts that take values from the standard base of the Hilbert space of dimension $d$.

\subsection{Weyl's adjoint operator}

One of the analyses conducted in this study involved establishing a correspondence between Weyl's adjoint operator,

 \begin{equation} \label{weylEquation}
 	W_{q,p}^\dagger = \sum_{k=0}^{d-1} \omega^{k\hspace{0.5pt}q}\Ket{k \oplus p}\Bra{k}
\end{equation}

and the $I, X, Y$, and $Z$ gates in high dimensions:

 \begin{equation}
  \begin{gathered}
    I^m = W_{0,0}^\dagger \\[1em]
    X^m = W_{0,m}^\dagger \\[1em]
    Z^m = W_{m,0}^\dagger \\[1em]
    Y^m = i^{(m\%d)} \; W_{m,m}^\dagger \\[1em]
   \end{gathered}
\end{equation}

In this manner, it is possible to find the $I, X, Y$, and $Z$ gates for any dimension raised to any exponent. The identity operator for high dimensions can be found using equation \eqref{weylEquation}  by replacing (q,p) with (0,0):

\begin{equation}
    \begin{gathered}
    I^m = W_{0,0}^\dagger \\[1em]
    = \sum_{k=0}^{d-1} \omega^0\Ket{k \oplus 0}\Bra{k} \\[1em]
    = \sum_{k=0}^{d-1} \Ket{k}\Bra{k} \\[1em]
    = \sum_{k=0}^{d-1} P_k \\[1em]
    \end{gathered}
\end{equation}

where $P_k$ is known as the projection operator, which is equal to $\Ket{k}\Bra{k}$.

By replacing (q,p) with (0,m) in equation \eqref{weylEquation}, it is possible to find the expression that generalizes the $X$ Pauli matrix for high dimensions and raising to any exponent:

\begin{equation}
    \begin{gathered}
    X^m = W_{0,m}^\dagger \\[1em]
    = \sum_{k=0}^{d-1} \omega^0\Ket{k \oplus m}\Bra{k} \\[1em]
    = \sum_{k=0}^{d-1} \Ket{k \oplus m}\Bra{k} \\[1em]
    \end{gathered}
\end{equation}

To find the generalized form of the $Z$ Pauli matrix, simply replace (q,p) with (m,0) in equation \eqref{weylEquation}:

\begin{equation}
    \begin{gathered}
    Z^m = W_{m,0}^\dagger = \sum_{k=0}^{d-1} \omega^{k\hspace{0.5pt}m}P_k \\[1em]
    = \sum_{k=0}^{d-1} \omega^{k\hspace{0.5pt}m}\Ket{k \oplus 0}\Bra{k} \\[1em]
    = \sum_{k=0}^{d-1} \omega^{k\hspace{0.5pt}m}\Ket{k}\Bra{k} \\[1em]
    = \sum_{k=0}^{d-1} \omega^{k\hspace{0.5pt}m}P_k \\[1em]
    \end{gathered}
\end{equation}

The $Y$ matrix can be viewed as a combination of the $X$ and $Z$ matrices. It is possible to find its generalized equation for high dimensions and raising to any exponent by using equation \eqref{weylEquation} again, replacing (q,p) with (m,m) and multiplying by a cyclical global phase of $i^{(m\%d)}$:

\begin{equation} \label{generalizedYPauli}
    \begin{gathered}
    Y^m = i^{(m\%d)} \; W_{m,m}^\dagger \\[1em]
    Y^m = i^{(m\%d)} \; \sum_{k=0}^{d-1} \omega^{k\hspace{0.5pt}m} \Ket{k \oplus m}\Bra{k}
    \end{gathered}
\end{equation}\break

In this manner, we find the $X^m$, $Y^m$, and $Z^m$ gates generalized to $d$ dimensions of the qubit-based $X$, $Y$, and $Z$ gates, respectively, raised to the $m$ power. Below is a summary of the gate generalizations described thus far:

\begin{figure}[H]
  \centering
  \includegraphics[width=14cm, height=4.5cm]{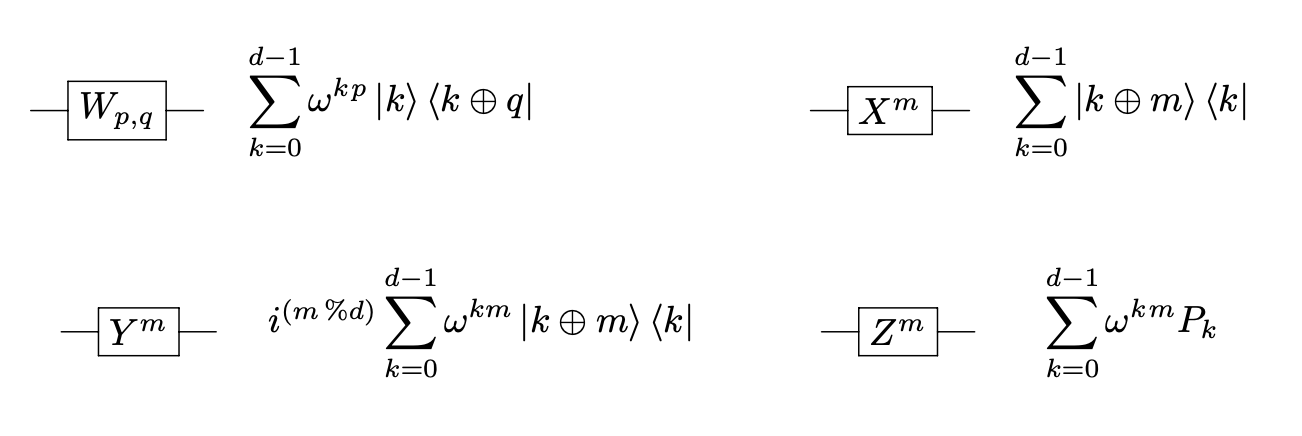}
  \caption{Single-qudit high-dimensional gates $X^m$, $Y^m$, $Z^m$, and $W_{p,q}$. \\
  $P_k$ is the projection operator equal to $\Ket{k}\Bra{k}$, and $\omega$ is defined by $e^{ (\frac{2\pi}{d}) i }$} 
\end{figure}

\subsection{Controlled gate}

The $CNOT$ gate, which acts on qubits, can be generalized to high dimensions by the $SUM$ gate, which is defined by the following expression:

\begin{equation}
SUM = \sum_{k=0}^{d-1} P_k \otimes X^k
\end{equation}

This gate can also be seen as a particular case of the arbitrary generalized controlled gate:

\begin{equation} \label{CUGateEquation}
CU = \sum_{k=0}^{d-1} P_k \otimes U^k
\end{equation}

where $U$ can be $X, Y, Z$, or any high-dimensional operator.

In this manner, the high-dimensional controlled gates available in QuantumSkynet are as follows:

\begin{figure}[H]
  \centering
  \includegraphics[width=13cm, height=3.5cm]{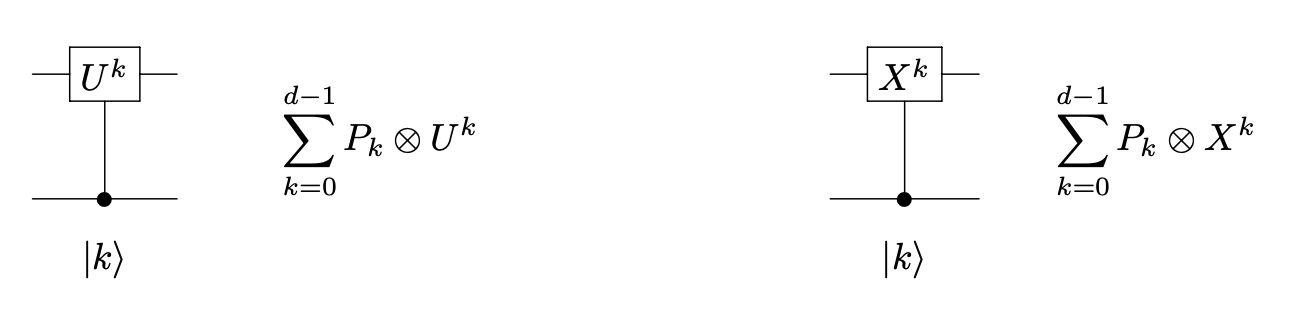}
  \caption{Multi-qudit generalized controlled gate and $SUM$ gate. \\
  U is an arbitrary unit matrix, and $P_k$ is the projection operator equal to $\Ket{k}\Bra{k}$} 
\end{figure}

\subsection{Generalized quantum Fourier transform}

The classic discrete Fourier transform maps a series of $N$ complex numbers into another series of complex numbers as follows \cite{camps2020quantum}:

\begin{equation}
    y_k = \frac{1}{\sqrt{N}} \sum_{j=0}^{N-1} \omega_N^{k\hspace{0.5pt}j} x_j \;, \quad k = 0, 1, ..., N-1
\end{equation}

where $x = [x_0, ..., x_N-1]^T  \in \mathbb{C}^N \; , \; y = [y_0, ..., y_N-1]^T  \in \mathbb{C}^N \; , \; \omega_N = e^{ (\frac{2\pi}{N}) i }$.

\vspace{3mm}
Similarly, the quantum Fourier transform acts on a qudit with quantum state of dimension $d$ (where $N=d$) and maps it to another $d$-dimensional quantum state according to the following expression:

\begin{equation}
    \Ket{j} \xrightarrow{\mathscr{F}} \frac{1}{\sqrt{d}} \sum_{k=0}^{d-1} \omega^{k j} \Ket{k}
\end{equation}

where $j$ is a quantum state belonging to the standard base of a $d$-dimensional Hilbert space with $\omega=e^{\frac{2 \pi i} {d}}$.

When $d$ is equal to 2, the quantum Fourier transform is the Hadamard gate acting on 1 or $N$ qubits. $d$ greater than 2 gives a generalized Hadamard gate that acts on 1 or $N$ qudits, which is known as $DFT$ in some papers \cite{lu2020quantum, imany2019high}.

We grouped all different combinations in which the quantum Fourier transform can be applied for any number of qudits and any number of dimensions, which we call a Generalized $QFT$ ($GQFT$).
For this, it is necessary to cover all elements of the quantum state resulting from the interaction between these qudits. That is, the size $N$ of the vector that will be affected by the $GQFT$ will be $d^n$, which is the same size as the tensor product between the qudits affected by the gate:

\begin{equation}
    \Ket{j} \xrightarrow{\mathscr{F}} \frac{1}{\sqrt{d^n}} \sum_{k=0}^{d^n-1} \omega^{k j} \Ket{k}
\end{equation}

where $n$ refers to the number of qudits of $d$ dimensions that will be affected by the gate.

The following figure shows the proposed $GQFT$.

\begin{figure}[H]
  \centering
  \includegraphics[width=7cm, height=7cm]{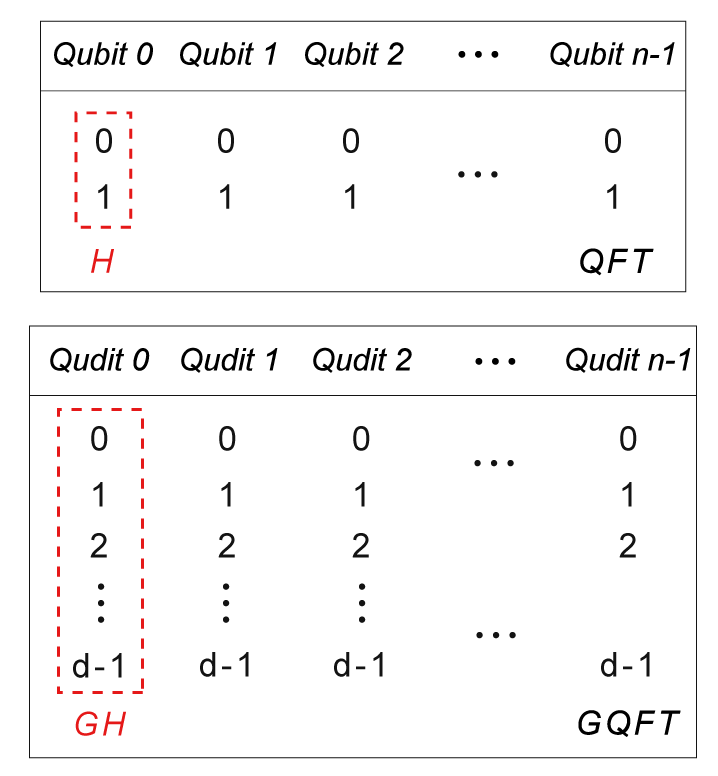}
  \caption{Proposed GQFT}
\end{figure}

As can be seen in the previous figure, the Hadamard gate for the case of qubits is a special case of the $QFT$ gate, while for the case of qudits, the $GH$ or Generalized Hadamard gate (commonly misnamed $DFT$) is a special case of the $GQFT$ gate.

Therefore, this is another gate available for use with QuantumSkynet:

\begin{figure}[H]
  \centering
  \includegraphics[width=8.5cm, height=2.5cm]{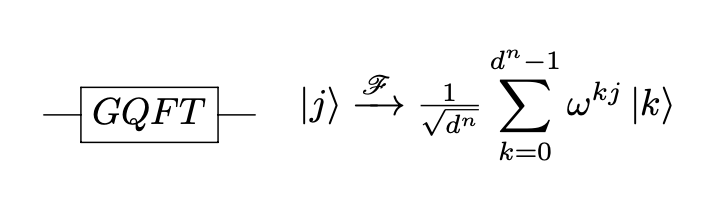}
  \caption{GQFT gate} 
\end{figure}

\section{Methods}

QuantumSkynet was written using the C++ programming language. It is divided into two core classes, Qudit and QRegister, as described below.

\subsection{Qudit class}
Consider the following definition of a qudit:
\begin{equation}
\Ket{\Psi} = \sum_{k=0}^{d-1} \alpha_k \Ket{k}
\end{equation}
where $d$ is the number of dimensions and $\alpha_k$ are the complex amplitudes that accompany the base states of the qudit.
A class for the construction of qudits was implemented in the simulator as follows:

\begin{codelisting}
#include <vector>
#include <complex>

class Qudit {
private:
  vector<complex<double>> amplitudes;

public:
  Qudit(vector<complex<double>> coefficients) {
    amplitudes = coefficients;
  };
  vector<complex<double>> getAmplitudes() {
    return amplitudes;
  };
};
\end{codelisting}

This class receives a vector of complex values that correspond to the amplitudes $\alpha_k$ that make up the quantum state of the qudit. This vector is assigned to the private attribute amplitudes, which can be accessed through the getAmplitudes() function.

\subsection{QRegister class}

To store and process the input qudits and allow their interaction between themselves and other functions in the quantum circuit, the QRegister class was created to define the structure of the quantum register:

\begin{codelisting}
class QRegister {
  private:
    int quditsDimensions;
    int quditsNumber;
    vector<Qudit> initialQudits;
    vector<complex<double>> statesVector;

  public:
    QRegister(
      int dimensions,
      vector<vector<complex<double>>> initialQuditsState
    ) {
      quditsDimensions = dimensions;
      quditsNumber = initialQuditsState.size();
      if (quditsNumber == 1) {
        Qudit qudit(initialQuditsState[0]);
        initialQudits.push_back(qudit);
        statesVector = initialQuditsState[0];
      } else {
        for (int i=0; i<quditsNumber; i++) {
          Qudit qudit(initialQuditsState[i]);
          initialQudits.push_back(qudit);
        }
        statesVector = tensorProduct(
            dimensions,
            initialQudits
        );
      }
      statesVector = normalize(statesVector);
    }

    int getDimensions() {
      return quditsDimensions;
    };

    int getQuditsNumber() {
      return quditsNumber;
    };

    vector<Qudit> getInitialQudits() {
      return initialQudits;
    };

    vector<complex<double>> getStatesVector() {
      return statesVector;
    };
};
\end{codelisting}

The $statesVector$ variable defined within the QRegister class stores the vector of amplitudes with the global quantum state of the circuit. This variable is modified by the application of the high-dimensional quantum gates supported by the simulator.

\subsection{Cloud-based architecture}

To allow access to the different functions of the simulator, cloud computing services provided by Amazon Web Services (AWS) were used. The following diagram shows the set of AWS-specific technologies that allowed the simulator to be deployed on the web:

\begin{figure}[H]
  \centering
  \includegraphics[width=10cm, height=6cm]{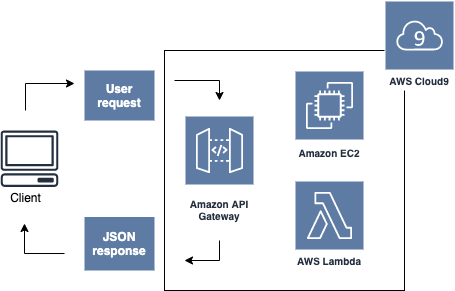}
  \caption{Cloud-based architecture}
\end{figure}

As shown in the previous diagram, this project was deployed in the form of an application programming interface (API) to be used as a gateway to the data and functionalities of the simulator.

A Cloud9 instance was used to debug the code and configure the simulator deployment on the cloud. Cloud9 automatically creates an EC2 instance and manages the lifecycle of this instance as required.

From the EC2 instance, a Lambda function was created in C++ with the logic to process the requests received by the client and return a response from the simulator. An API was then created using the Amazon API Gateway service and integrated with the pre-built Lambda function.

To request information from the API, an HTTP request is sent from the client with the quantum circuit to be simulated in JSON format in the body of the request, and a response that also arrives in JSON format is expected.

\section{Results and discussion}

\subsection{Quantum phase estimation algorithm}
Quantum phase estimation is one of the most important subroutines in quantum computation \cite{paesani2017experimental}. This algorithm is a central part of many other quantum algorithms, including the Shor factoring algorithm.

The high-dimensional version of this algorithm, which was presented in \cite{lu2020quantum}, was studied and implemented in QuantumSkynet. The quantum circuit associated with the phase estimation algorithm (PEA) is shown below.

\begin{figure}[H]
  \centering
  \includegraphics[width=14cm, height=5cm]{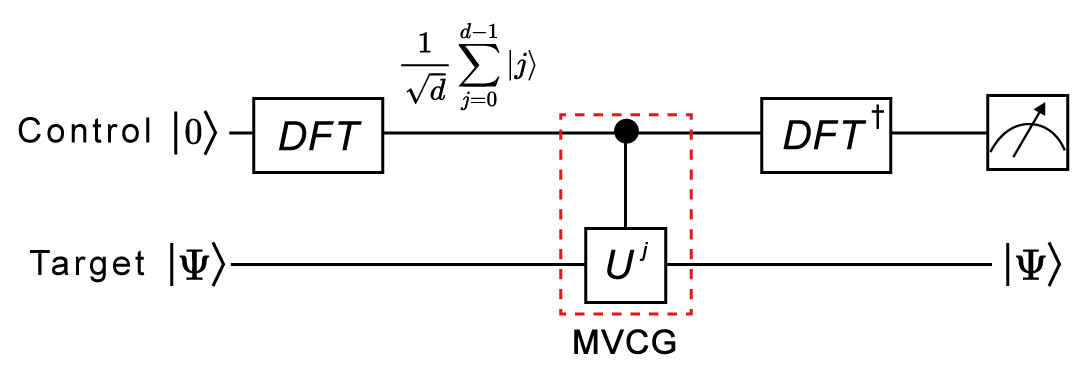}
  \caption{Phase estimation algorithm}
\end{figure}

The $DFT$ gates are qudit Hadamard gates, which can be represented by the $GQFT$ gate for a qudit in QuantumSkynet, and the $U$ gate used to test this algorithm was the same as that was used as a proof of concept in \cite{lu2020quantum}, the generalized 3-dimensional $Z$ gate.

The eigen-phases that correspond to each of the eigen-states of the generalized gate $Z$ are as follows:

\begin{center}
 \begin{tabular}{||c c c c||} 
 EigenState & $\Ket{0}$ & $\Ket{1}$ & $\Ket{2}$ \\ [2ex] 
 True Phase, $\phi$ & 0 & $\frac{2\pi}{3}$ & $\frac{4\pi}{3}$ \\  [1ex] 
\end{tabular}
\end{center}

The simulation of the above circuit in QuantumSkynet was conducted as follows:

\begin{figure}[H]
  \centering
  \includegraphics[width=11cm, height=3.8cm]{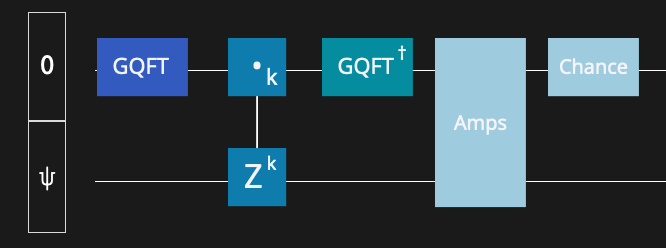}
  \caption{PEA in QuantumSkynet}
\end{figure}

where $\Ket{\Psi}$ represents each of the eigen-states $\Ket{0}, \Ket{1} y \Ket{2}$ required to find the respective eigen-phases.

Below are the respective resulting states based on the vector of amplitudes and value $\tau$ equal to the measured qudit value with index 0 based on the vector of probabilities, calculated by QuantumSkynet for each eigen-state (the qubit at the bottom of the circuit is the least significant):

\begin{center}
 \begin{tabular}{||c c c c||} 
 EigenState & $\Ket{0}$ & $\Ket{1}$ & $\Ket{2}$ \\ [2ex] 
 Result & $\Ket{00}$ & $\Ket{11}$ & $\Ket{22}$ \\  [2ex] 
 $\tau$ & 0 & 1 & 2 \\  [1ex] 
\end{tabular}
\end{center}

The phase $\phi$ is equal to $2\pi\tau/3$. Therefore, the values of the phases found by QuantumSkynet for the respective eigen-states of the 3-dimensional $Z$ matrix were as follows:

\begin{center}
 \begin{tabular}{||c c c c||} 
 EigenState & $\Ket{0}$ & $\Ket{1}$ & $\Ket{2}$ \\ [2ex] 
 True Phase, $\phi$ & 0 & $\frac{2\pi}{3}$ & $\frac{4\pi}{3}$ \\  [1ex] 
\end{tabular}
\end{center}

which are effectively the values of the phases presented in \cite{lu2020quantum}.

\subsection{Deutsch--Jozsa algorithm}
The Deutsch--Jozsa algorithm is a remarkable example that demonstrates quantum speed-up. This algorithm can determine whether a function $f(x)$ is constant or balanced using only one function evaluation.

\begin{figure}[H]
  \centering
  \includegraphics[width=15cm, height=5cm]{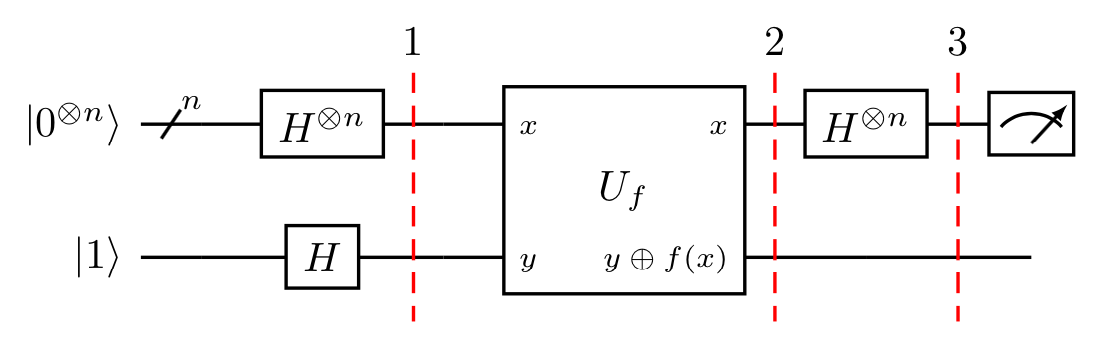}
  \caption{Deutsch--Jozsa algorithm}
\end{figure}

In the case of qudits, the Hadamard gate is replaced by the $GQFT$ gate, but the analysis to identify a function as constant or balanced is similar to that for the two-dimensional case; if each x-register qudit returns 0, it is a constant function; otherwise, the function is balanced \cite{wang2020qudits}.

As an example, the following high-dimensional quantum circuits were tested on QuantumSkynet to evaluate whether the functions $f(x)$ are constant or balanced:

\begin{figure}[H]
  \centering
  \includegraphics[width=15cm, height=6cm]{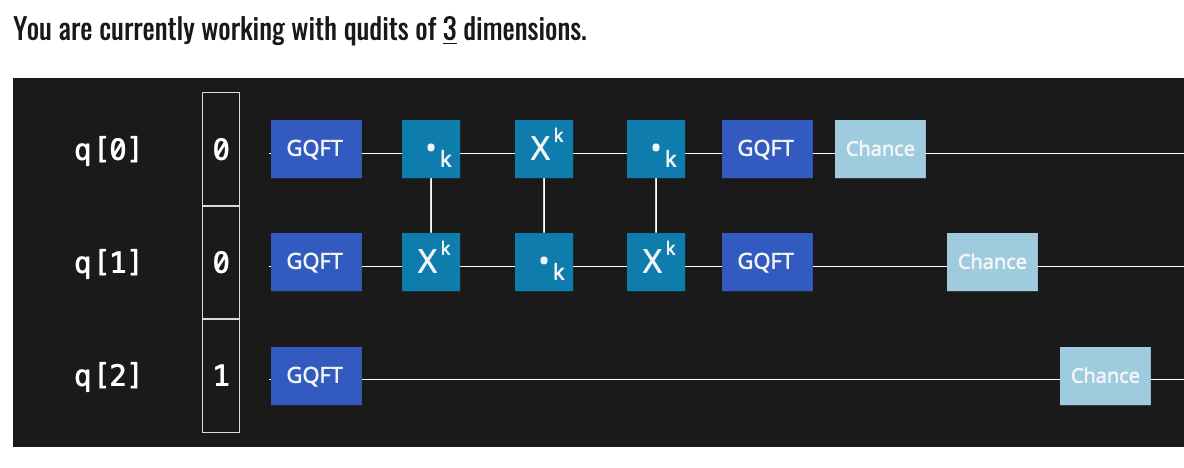}
  \caption{Deutsch--Jozsa algorithm on QuantumSkynet}
\end{figure}

For the previous circuit, the following probability vector was obtained for each qudit:

\begin{center}
 \begin{tabular}{c c c c}
 \multicolumn{1}{c}{} & 
 \multicolumn{3}{c}{Chance result} \\[1ex] \hline
  & $\Ket{0}$ & $\Ket{1}$ & $\Ket{2}$ \\ [1ex] \hline
 q[0] & 1 & 0 & 0 \\  [1ex] 
 q[1] & 1 & 0 & 0 \\  [1ex] 
 q[2] & 0.33 & 0.33 & 0.33 \\  [1ex] 
\end{tabular}
\end{center}

The probability of obtaining $\Ket{0}$ for the qudits at the top of the circuit was $100\%$, which means that the function is constant. A similar occurrence is seen with the following circuit, where all the probability values for the two least significant qudits of the circuit are equal to 1:

\begin{figure}[H]
  \centering
  \includegraphics[width=15cm, height=7cm]{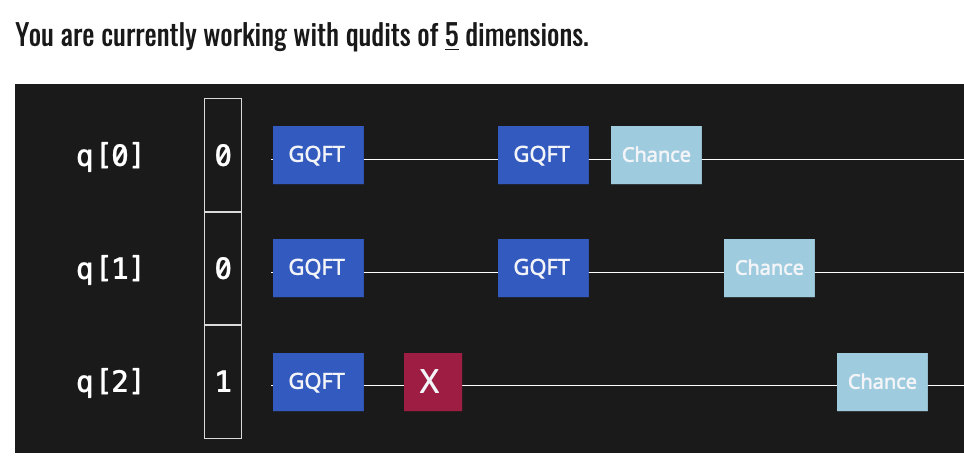}
  \caption{Deutsch--Jozsa algorithm on QuantumSkynet}
\end{figure}

\begin{center}
 \begin{tabular}{c c c c c c} 
 \multicolumn{1}{c}{} & 
 \multicolumn{5}{c}{Chance result} \\[1ex] \hline
  & $\Ket{0}$ & $\Ket{1}$ & $\Ket{2}$ & $\Ket{3}$ & $\Ket{4}$ \\ [1ex] \hline
 q[0] & 1 & 0 & 0 & 0 & 0 \\  [1ex] 
 q[1] & 1 & 0 & 0 & 0 & 0 \\  [1ex] 
 q[2] & 0.2 & 0.2 & 0.2 & 0.2 & 0.2 \\  [1ex] 
\end{tabular}
\end{center}

Now we show two examples of quantum circuits in which the function $f(x)$ is balanced:

\begin{figure}[H]
  \centering
  \includegraphics[width=15cm, height=7cm]{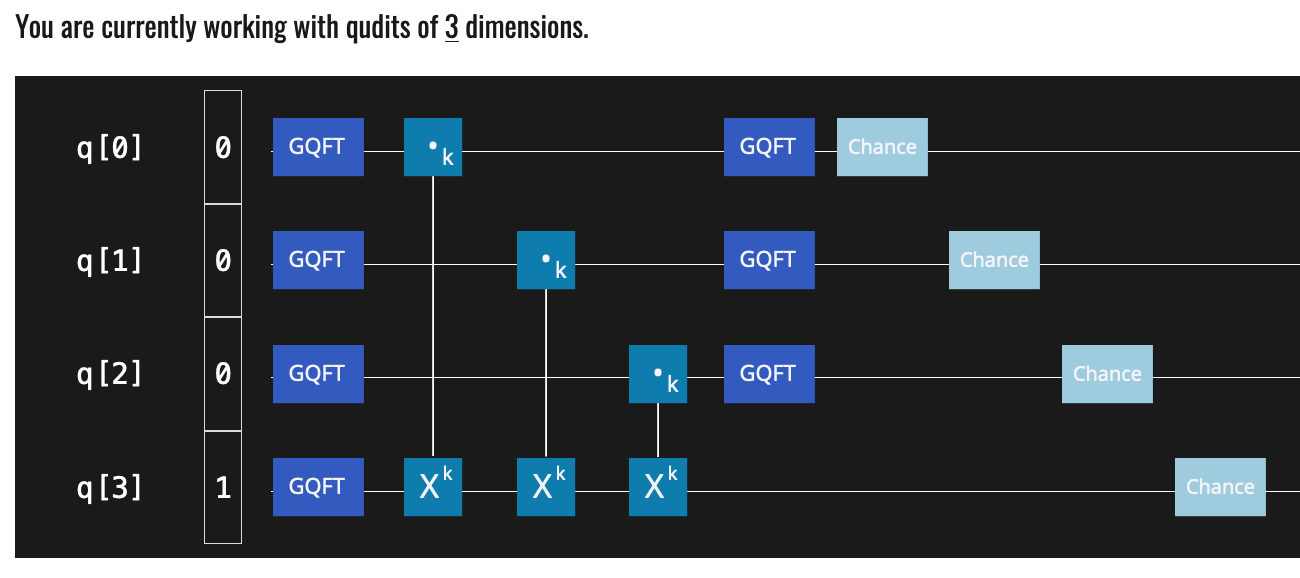}
  \caption{Deutsch--Jozsa algorithm on QuantumSkynet}
\end{figure}

\begin{center}
 \begin{tabular}{c c c c} 
 \multicolumn{1}{c}{} & 
 \multicolumn{3}{c}{Chance result} \\[1ex] \hline
  & $\Ket{0}$ & $\Ket{1}$ & $\Ket{2}$ \\[1ex] \hline 
 q[0] & 0 & 1 & 0  \\  [1ex] 
 q[1] & 0 & 1 & 0 \\  [1ex] 
 q[2] & 0 & 1 & 0 \\  [1ex]
 q[3] & 0.33 & 0.33 & 0.33 \\  [1ex]
\end{tabular}
\end{center}

Because the three least significant qudits have zero probability of being at $\Ket{0}$, the function $f(x)$ is balanced.

\begin{figure}[H]
  \centering
  \includegraphics[width=15cm, height=5cm]{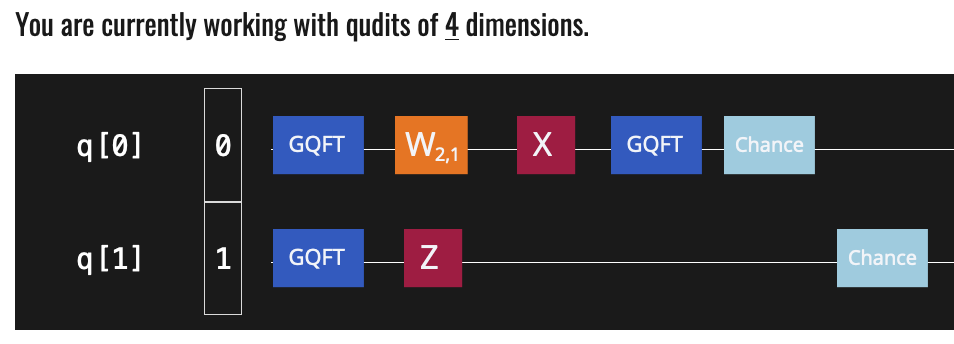}
  \caption{Deutsch--Jozsa algorithm on QuantumSkynet}
\end{figure}

\begin{center}
 \begin{tabular}{c c c c c } 
 \multicolumn{1}{c}{} & 
 \multicolumn{4}{c}{Chance result} \\[1ex] \hline
  & $\Ket{0}$ & $\Ket{1}$ & $\Ket{2}$ & $\Ket{3}$ \\[1ex] \hline 
 q[0] & 0 & 0 & 1 & 0  \\  [1ex] 
 q[1] & 0.25 & 0.25 & 0.25 & 0.25 \\  [1ex] 
\end{tabular}
\end{center}

In this last example, the function $f(x)$ is also balanced because the least significant qudit has no chance of being $\Ket{0}$.

\section{Conclusion}

Due to the advantages of qudit-based circuits over qubit-based ones, high-dimensional quantum computing simulation will continue to be an area of high interest.

The calculations derived from the analysis of quantum circuits based on qudits can become complex and grow as the number of dimensions increases. In this study, we explored a very useful tool to easily and intuitively simulate high-dimensional quantum circuits.

QuantumSkynet opens to the possibility of helping extend qubit-based quantum circuits to high-dimensional circuits as well as exploring and analyzing new qudit-based quantum circuits.

As future work, we intend to add support for the distinction between pure and mixed states and allow the simulation of different levels of quantum decoherence.

\printbibliography[title={References}]
\end{document}